\documentclass[11pt]{article}
\usepackage[a4paper,margin=25mm]{geometry}
\usepackage{amsmath,amssymb,amsthm}
\usepackage[T1]{fontenc}
\usepackage{lmodern,microtype}
\usepackage{graphicx}
\usepackage{caption}
\usepackage{float}
\usepackage{placeins}
\usepackage{xurl}
\usepackage[hidelinks]{hyperref}
\newtheorem{lemma}{Lemma}[section]
\newtheorem{proposition}[lemma]{Proposition}
\newtheorem{theorem}[lemma]{Theorem}
\newtheorem{corollary}[lemma]{Corollary}
\numberwithin{equation}{section}
\newcommand{\eps}{\epsilon}
\newcommand{\DD}{\Delta K}
\newcommand{\Tr}{\mathcal T_\eps}
\newcommand{\yr}{y_{\mathrm{rough}}}
\newcommand{\gr}{\gamma_{\mathrm{rough}}}
\newcommand{\Dr}{\Delta K_{\mathrm{rough}}}
\newcommand{\ee}{\mathfrak e}
\newcommand{\Logder}{\partial_{\log x}}
\newcommand{\ZenodoDOI}{\href{https://doi.org/10.5281/zenodo.23776799}{\nolinkurl{10.5281/zenodo.23776799}}}
\hypersetup{pdftitle={Ultraviolet regularization of singular switching optima in an Ising--Kuramoto model},pdfauthor={Hayashi Reon}}
\title{Ultraviolet regularization of singular switching optima in an Ising--Kuramoto model}
\author{Hayashi Reon\\[0.4em]
\small Hiroshima Prefectural Fukuyama Seishikan Senior High School,\\
\small Fukuyama, Hiroshima, Japan\\[0.3em]
\small Corresponding author: \texttt{reon.0627.che@gmail.com}}
\date{}
\begin{document}
\maketitle
\begin{abstract}
We study how finite ultraviolet bandwidth regularizes optimal switching in the mean-field linear-instability threshold of a dynamic Ising--Kuramoto model. For a specified ideal rough-noise family whose free-phase correlation obeys \(1-G(t)\propto t^{1+\epsilon}\), \(0<\epsilon<1\), the selected switching rate diverges essentially as \(\gamma_{\rm rough}\tau\sim e\,r^{1/\epsilon}\) when \(\epsilon\downarrow0\), where \(\tau\) is the reference correlation time and \(r>1\) is a fixed noise-to-response ratio. For a Lorentzian spectral cutoff of frequency \(\Lambda\), we control the true full-susceptibility global optimizer through the singular crossover between rough, cutoff-dominated, and Lambert-\(W\) regimes, including relative control of the vanishing threshold gain. Writing \(y_{\rm rough}=\gamma_{\rm rough}/\Lambda\), we prove, within the stated bounded inner paths, that recovery of the ideal optimal rate is equivalent to \(y_{\rm rough}/\sqrt{\epsilon}\to0\), whereas recovery of the ideal threshold gain is equivalent to the stronger condition \(y_{\rm rough}/\epsilon\to0\). Thus the optimizer can agree asymptotically with the ideal model while the achieved gain remains ultraviolet-sensitive. A Gaussian filtered-white endpoint comparison further shows that the limiting scaled optimizer and gain coefficients depend on the cutoff convention. These results quantify a noncommuting interplay between optimization and short-time regularization in an effective stochastic description.
\end{abstract}

\begingroup
\clubpenalty=10000
\widowpenalty=10000
\displaywidowpenalty=10000
\interlinepenalty=10000
\section{Introduction}
Collective synchronization in systems with switching internal states depends on how switching dynamics interact with phase correlations \cite{hayashi2026}. Coupled phase oscillators provide a standard framework for synchronization \cite{acebron2005}; colored forcing modifies its onset \cite{tonjes2010}, and intermediate rates of prescribed network switching can favor synchronization \cite{zhang2021}. In the internal-switching model considered here, changing the switching rate modifies both the response to a collective field and the time over which that response remains coherent. The issue considered here is how an optimal rate selected by an ideal noise description survives when the shortest correlation scales are regularized.

The preceding study of a dynamic Ising--Kuramoto model \cite{hayashi2026} expressed the mean-field instability threshold as
\begin{equation}
K_c(\gamma)=\frac{1}{(\mu+\beta_s\gamma)\chi_\gamma},\qquad
\chi_\gamma=\frac12\int_0^\infty e^{-\gamma t}G(t)\,dt,
\end{equation}
where \(G(t)\) is the free-phase correlation. This form separates the increasing response factor from the susceptibility sampled by switching. For Ornstein--Uhlenbeck forcing, the previous work identified finite optimal switching rates in the parameter regime studied there. That result provides the starting point for the present analysis.

The previous criterion already distinguishes integrated phase-memory moments from short-time behavior \cite{hayashi2026}. What remains unresolved is how ultraviolet regularization affects an optimizer that becomes singularly fast. At large \(\gamma\), the Laplace factor \(e^{-\gamma t}\) increasingly probes the behavior of \(G(t)\) near the origin. The ideal noise covariance is the exponentially tempered power-law kernel of Ref. \cite{molina2018}. For the specified ideal rough-noise family with \(1-G(t)\propto t^{1+\epsilon}\), the ideal optimizing rate has the essential singularity \(\gamma_{\rm rough}\tau\sim e\,r^{1/\epsilon}\) as \(\epsilon\downarrow0\), at fixed \(r=\beta_sD/\mu>1\). Here \(D\) is the integrated noise intensity and \(\tau\) is the reference correlation scale. The ideal forcing has a divergent equal-time covariance for \(0<\epsilon<1\). Related smooth Gaussian processes have been constructed using cascaded linear filters \cite{viggiano2020}. The finite-bandwidth regularization considered here restores the quadratic short-time behavior \(1-G(t)\sim C(0)t^2/2\), precisely where a sufficiently fast optimizer probes the noise.

More broadly, stochastic singular limits with more than one fast relaxation scale can depend on how those scales approach zero; rigorous white-noise limits provide examples in which the limiting equation changes with the relative rate of the fast scales \cite{pavliotis2005}. The present problem is different in a specific way: we do not seek only a limiting stochastic equation at a fixed observable scale. Optimization drives the selected switching rate itself toward the ultraviolet sector, and the quantity of interest includes a threshold gain that vanishes in the same limit. The question is therefore one of \emph{relative recovery}: the optimizer position may already agree with the ideal description while the small optimized gain remains sensitive to the regularization.

This raises a precise question: how does a finite ultraviolet scale regularize an optimizer that diverges essentially in the ideal rough limit? Answering it requires tracking both the position of the threshold minimum and its depth below the fast-switching threshold. These two quantities need not recover the ideal description on the same scale, even though they characterize the same optimum.

We establish four results. First, we obtain a controlled crossover for the full-susceptibility global optimizer under Lorentzian ultraviolet regularization, including relative control of the minimum depth. Second, we resolve rough and cutoff outer regions joined by a Lambert-\(W\) inner layer. Third, within the bounded inner scaling paths specified below, recovery of the rate and depth is characterized by the distinct conditions \(y_{\rm rough}/\sqrt\epsilon\to0\) and \(y_{\rm rough}/\epsilon\to0\), respectively, where \(y_{\rm rough}=\gamma_{\rm rough}/\Lambda\). Fourth, comparison with the Gaussian filtered-white endpoint shows that the limiting scaled optimizer and depth coefficients depend on the cutoff convention.

The analysis concerns a mean-field linear instability threshold. It does not describe the nonlinear ordered phase, establish finite-size observability, or provide an experimental realization of the noise regularization.
\section{Model and setup}
In the underlying globally coupled model, each unit carries an Ising spin \(\sigma_j=\pm1\) and a phase \(\theta_j\). The instability considered here is the spin-weighted first harmonic \(Z_m=N^{-1}\sum_j\sigma_j e^{i\theta_j}\) of the nonmagnetic incoherent state, with no direct phase-only coupling. In this first-harmonic sector, the microscopic dynamics relevant at linear order can be written
\begin{equation}
\dot\theta_i=\mu K\sigma_i\,\operatorname{Im}(Z_m e^{-i\theta_i})+\eta_i,
\qquad
w_i(\sigma_i\!\to\!-\sigma_i)=\frac{\gamma}{2}
\left[1-\sigma_i\tanh(\beta_s h_i)\right],
\quad
h_i=K\operatorname{Re}(Z_m e^{-i\theta_i}).
\end{equation}
The phase forcings are independent across units, with
\begin{equation}
\langle\eta_i(t)\eta_j(t')\rangle=\delta_{ij}C(t-t'),
\end{equation}
and, in the noninteracting incoherent reference state, the free spin process is statistically independent of the phase forcing.
Terms proportional to the ordinary magnetization decouple from this angular mode at linear order about the nonmagnetic background, and a direct phase-only coupling is set to zero. At zero local field the free spin has correlation \(\langle\sigma(t)\sigma(0)\rangle=e^{-\gamma t}\). Linear response of \(Z_m\) then has two additive channels: the phase drift contributes \(\mu\), while the Glauber spin response contributes \(\beta_s\gamma\); both sample the free-phase correlation through the same exponential damping. At marginality their self-consistency condition is
\begin{equation}
1=K(\mu+\beta_s\gamma)\frac12\int_0^\infty e^{-\gamma t}G(t)\,dt.
\end{equation}
Here \(\mu>0\) is the phase mobility, \(\beta_s>0\) is the spin-response sensitivity, and \(\gamma\ge0\) is the zero-field spin-relaxation rate. The full linear-response derivation is given in Ref. \cite{hayashi2026}. With \(s=\mu/\beta_s\), the threshold is therefore equivalently
\begin{equation}
H(\gamma)=(\gamma+s)\int_0^\infty e^{-\gamma t}G(t)\,dt,
\qquad K_c(\gamma)=\frac{2}{\beta_sH(\gamma)}.
\end{equation}

Let \(\eta(t)\) be centered stationary Gaussian phase forcing with covariance \(C(t)\). Its integrated phase increment gives the exact free correlation
\begin{equation}
G(t)=\left\langle\exp\!\left[i\int_0^t\eta(u)\,du\right]\right\rangle
=e^{-Q(t)},\qquad
Q(t)=\int_0^t(t-u)C(u)\,du.
\end{equation}
We use the exponentially tempered power-law covariance studied in Ref. \cite{molina2018}, with its intensity normalized here by its time integral:
\begin{equation}
C_\epsilon(t)=\frac{D}{\tau^\epsilon\Gamma(\epsilon)}
|t|^{\epsilon-1}e^{-|t|/\tau},\qquad 0<\epsilon\le1.
\end{equation}
For \(0<\epsilon<1\), this is understood as the covariance kernel of generalized Gaussian forcing; its time integral defines the phase process. The parameter \(\epsilon\) controls the short-time singularity, while \(\tau\) sets the exponential envelope of the ideal covariance.

Using the Fourier convention \(S(\omega)=\int_{\mathbb R}C(t)e^{-i\omega t}\,dt\), we impose the Lorentzian spectral regularization
\begin{equation}
S_{\epsilon,\Lambda}(\omega)=
\frac{2D\,\operatorname{Re}(1+i\omega\tau)^{-\epsilon}}
{1+\omega^2/\Lambda^2}.
\end{equation}
The frequency \(\Lambda>0\) sets the ultraviolet filter scale. Here ``finite bandwidth'' refers to a characteristic ultraviolet spectral roll-off scale rather than compact spectral support. The regularization preserves \(S(0)=2D\), hence \(\int_0^\infty C(t)\,dt=D\), while making \(C(0)\) finite. Consequently, the regularization restores the quadratic short-time behavior \(1-G(t)\sim C(0)t^2/2\) as \(t\downarrow0\). The ideal kernel and the Lorentzian filter specify the noise description; the results below concern the optimization induced by this choice.

The dimensionless parameters organizing the analysis are
\begin{equation}
r=\frac{D}{s},\qquad L=\Lambda\tau,\qquad
z=\epsilon\log L-\log r.
\end{equation}
Here \(r\) compares the integrated noise intensity with the response scale \(s\), and \(L\) compares the filter frequency with \(\tau^{-1}\). The coordinate \(z\) measures their competition on the logarithmic scale relevant to the exponentially large ideal optimizer. We hold \(D,\tau,\mu,\beta_s>0\) fixed with \(r>1\), and consider \(\epsilon\downarrow0\), \(L\to\infty\). The inner paths are \(z=\epsilon\log(1/\epsilon)\xi\) with \(|\xi|\le M\) for fixed finite \(M\); the necessary and sufficient recovery conditions are restricted to this path class.

For reference, the singular ideal-rough scale can be obtained without differentiating an unspecified little-\(o\) remainder. For the unregularized covariance, its Laplace transform is exactly
\begin{equation}
B_{\rm ideal}(\gamma)=D(1+\gamma\tau)^{-\epsilon}.
\end{equation}
The derivative-controlled large-rate reduction used in the Appendix gives the stationarity condition
\begin{equation}
s\sim B_{\rm ideal}(\gamma)-\gamma B'_{\rm ideal}(\gamma)
\sim D(1+\epsilon)(\gamma\tau)^{-\epsilon}.
\end{equation}
Hence
\begin{equation}
\gamma_{\rm rough}\tau\sim[(1+\epsilon)r]^{1/\epsilon}
\sim e\,r^{1/\epsilon},
\end{equation}
because \((1+\epsilon)^{1/\epsilon}\to e\). The same controlled expansion gives \(-B'_{\rm ideal}(\gamma_{\rm rough})\sim s\epsilon/[(1+\epsilon)\gamma_{\rm rough}]\), and therefore the associated ideal depth is
\begin{equation}
\Delta K_{\rm rough}\sim
\frac{2s\epsilon}{\beta_s(1+\epsilon)\gamma_{\rm rough}}.
\end{equation}
These formulas define the ideal reference scales used below; the Appendix supplies the uniform error control needed when the relevant slopes and depths shrink with \(\epsilon\).

We denote the global maximizing rate of \(H\), when interior, by \(\gamma_*\). Equivalently, it minimizes \(K_c\) over \(\gamma\ge0\). Since the fast-switching threshold is \(2/\beta_s\), its depth is
\begin{equation}
\Delta K=\frac{2}{\beta_s}-K_c(\gamma_*).
\end{equation}
All optimization statements refer to the full correlation \(G=e^{-Q}\). The covariance reductions used to analyze this optimum, their error control, and the existence and uniqueness statements are given in the Appendix.
\section{Results}
\subsection{Finite-bandwidth minima}

All numerical examples use \(D=\tau=\mu=1\) and \(\beta_s=2\), so that \(s=\mu/\beta_s=1/2\) and \(r=D/s=2\).

The numerical records used in Figs.~1--4 are archived in the associated Zenodo reproducibility record (DOI: \ZenodoDOI). The archive \nolinkurl{ESM_1_full_susceptibility_validation.zip} contains the fixed full-susceptibility validation runs, raw records, convergence checks, and scripts for regenerating the published validation tables. The archive \nolinkurl{ESM_2_publication_figures.zip} contains the exact figure data, plotting scripts, source records, numerical settings, and the additional cutoff-outer validation point. The included README files give the rerun commands and distinguish full-susceptibility calculations from covariance reductions and asymptotic curves. Replotting the figures does not require rerunning the optimizations.

Figure 1 shows that an interior threshold minimum persists in the representative Lorentzian-cutoff cases. Along the illustrated sequence, \(L=\Lambda\tau=2^{1/\epsilon}\), decreasing \(\epsilon\) moves the minimum to much larger switching rates while rapidly reducing its depth. Both the roughness exponent and bandwidth vary along this sequence. Plotting the threshold reduction exposes minima that become difficult to distinguish in the raw threshold, while rescaling by each full-susceptibility optimum resolves their local profiles. A well-defined minimizing rate can therefore coexist with a very small threshold gain.

\begin{figure}[H]
\centering
\includegraphics[width=\textwidth]{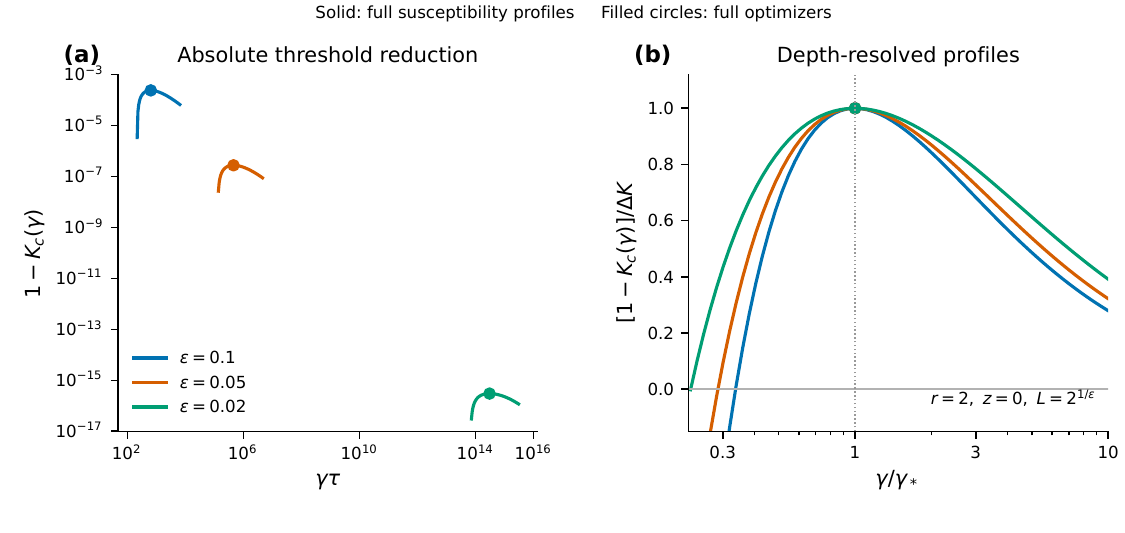}
\caption{Finite-cutoff threshold minima resolved by their depth. The Lorentzian-filtered model is shown at three validation-set parameter pairs, with \(\epsilon\) = 0.1, 0.05, and 0.02 along the fixed crossover coordinate \(z\) = 0; thus \(L=2^{1/\epsilon}\) and both the roughness exponent and bandwidth change between curves. Panel (a) plots the positive threshold reduction, \(1-K_c(\gamma)\), against \(\gamma\tau\) on logarithmic axes. Its maximum is the threshold minimum. Panel (b) rescales each horizontal axis by its own full-susceptibility optimizer and each vertical axis by its own minimum depth, exposing profiles that would be nearly invisible in a plot of the raw threshold. Solid curves are direct full-susceptibility evaluations on the archived 65-point relative-rate grid for each parameter pair; they are stored at double precision for display. Filled circles denote the multiprecision full-susceptibility optimizers. No covariance reduction or asymptotic profile is substituted for these curves. Negative threshold reductions are omitted from the logarithmic panel but retained within the displayed range of the normalized panel. This figure introduces the finite-bandwidth minimum and its changing position and depth; it is not an isolated-bandwidth comparison or a quantitative test of an outer asymptotic formula}
\label{fig:1}
\end{figure}
\subsection{Singular crossover of the optimizer}

Theorem~\ref{main} identifies three asymptotic regions for the full-susceptibility optimizer, shown together in Fig. 2.

In the rough outer region, at fixed \(z>0\),
\begin{equation}
\gamma_*\tau\sim[(1+\epsilon)r]^{1/\epsilon}.
\end{equation}
The optimum lies well below the cutoff rate and recovers the ideal rough prediction. The threshold depth also recovers its corresponding rough-model asymptote, defined below.

In the cutoff outer region, at fixed \(-\log r<z<0\), the optimum instead scales with the bandwidth:
\begin{equation}
\begin{gathered}\frac{\gamma_*}{\Lambda}\to y_0(z)
=e^{-z}-1+\sqrt{e^{-z}(e^{-z}-1)},\\
\Delta K\sim\frac{2s}{\beta_s\Lambda[1+2y_0(z)]}.\end{gathered}
\end{equation}
Thus a finite ultraviolet scale changes both the rate selected by the ideal model and the associated gain. The outer asymptotics are uniform on compact subsets of their respective open intervals.

The crossover is resolved on paths \(z=\epsilon\log(1/\epsilon)\xi\), with \(|\xi|\le M\) for any fixed finite \(M\). With the positive inner scale defined by the principal Lambert branch,
\begin{equation}
\begin{gathered}\lambda^2=\frac{\epsilon}{2}W_0\!\left[\frac{2}{\epsilon}e^{2-2z/\epsilon}\right],
\\
\frac{\gamma_*}{\Lambda\lambda}\to1,\qquad
\Delta K\sim\frac{2s}{\beta_s\Lambda}\left(1+\frac{\epsilon}{\lambda}\right),\end{gathered}
\end{equation}
both relative asymptotics hold uniformly over the stated bounded-\(\xi\) layer. The Appendix establishes these results for the true global optimum and controls the depth error relative to the depth itself.

The numerical comparison distinguishes three levels of description. Full-susceptibility points solve \(H'(\gamma_*)=0\) using the exact phase correlation in \(H(\gamma)=(\gamma+s)\int_0^\infty e^{-\gamma t}G(t)\,dt\). The finite-\(L\) covariance root \(\rho\) solves \(B(\rho)-\rho B'(\rho)=s\), where \(B\) is the exact Laplace transform of the regularized covariance. The dimensionless high-\(L\) root \(\widehat y\) solves \(\mathcal T_\epsilon(\widehat y)=e^z\), with \(\mathcal T_\epsilon\) defined in the Appendix. Exactness of the covariance transform does not make \(\rho\) identical to the full optimizer. Figure 2 labels these objects separately; their asymptotic equivalence is justified analytically.

The full points at \(z=0.3\) and \(z=-0.3\) approach the respective outer predictions as \(\epsilon\) decreases. The inner comparison remains less accurate at the available finite values: at \(z=0\) and \(\epsilon=0.02\), \(\gamma_*/(\Lambda\lambda)\simeq1.22\). The continuation to smaller \(\epsilon\) in Fig. 2(b) consists only of reduced-equation results. These comparisons are finite-\(\epsilon\) consistency checks; the analytic limits and their uniformity follow from Theorem~\ref{main}.

Full-susceptibility roots in the thirteen-point validation set were solved in \(\log\gamma\), using \(u=\gamma t\), the exact Lorentzian phase exponent, and \texttt{expm1} to evaluate small depths without subtracting nearly equal thresholds. The baseline used 50-digit arithmetic and \(u\le160\). All thirteen points were reevaluated with refined quadrature and \(u\le190\); four representative points from the twelve-point set and the added cutoff-outer point were also re-solved at 80 digits with \(u\le220\) and an equivalent special-function representation of \(Q\). For this set, the finite-\(L\) and high-\(L\) roots were computed separately at 80 and 110 digits. The archived convergence comparisons are summarized in Table~\ref{tab:conv}.
\begin{table}[H]
\centering
\small
\begin{tabular}{lc}
\hline
check & maximum relative change \\
\hline
12-point refined quadrature, \(\Delta K\) & 0 at 45-digit storage precision \\
representative full roots, 50\(\to\)80 digits: \(\gamma_*\) & \(2.15\times10^{-45}\) \\
representative full roots, 50\(\to\)80 digits: \(\Delta K\) & \(1.43\times10^{-45}\) \\
added cutoff-outer full root, 50\(\to\)80 digits: \(\gamma_*\) & \(8.20\times10^{-46}\) \\
added cutoff-outer full root, 50\(\to\)80 digits: \(\Delta K\) & \(7.77\times10^{-46}\) \\
reduced roots, 80\(\to\)110 digits & \(2.11\times10^{-68}\) \\
\hline
\end{tabular}
\caption{Numerical convergence checks recorded in the supplied reproducibility archives. The first entry means equality after storage to 45 significant digits, not exact mathematical equality. The 50-to-80-digit entries also change the integration cutoff from \(u=160\) to \(220\) and use an equivalent special-function representation of \(Q\). These are convergence checks rather than interval-arithmetic certification.}
\label{tab:conv}
\end{table}
The Lorentzian validation points, Gaussian endpoint points, and display profiles retain their separately documented precision settings in the accompanying reproducibility archives described above. In Fig.~1 the multiprecision optimizers and depths are computed before export; the 65-point profile arrays are stored at double precision only for plotting, so the tiny depth is not reconstructed by subtracting two nearly equal double-precision thresholds.

\begin{figure}[H]
\centering
\includegraphics[width=\textwidth]{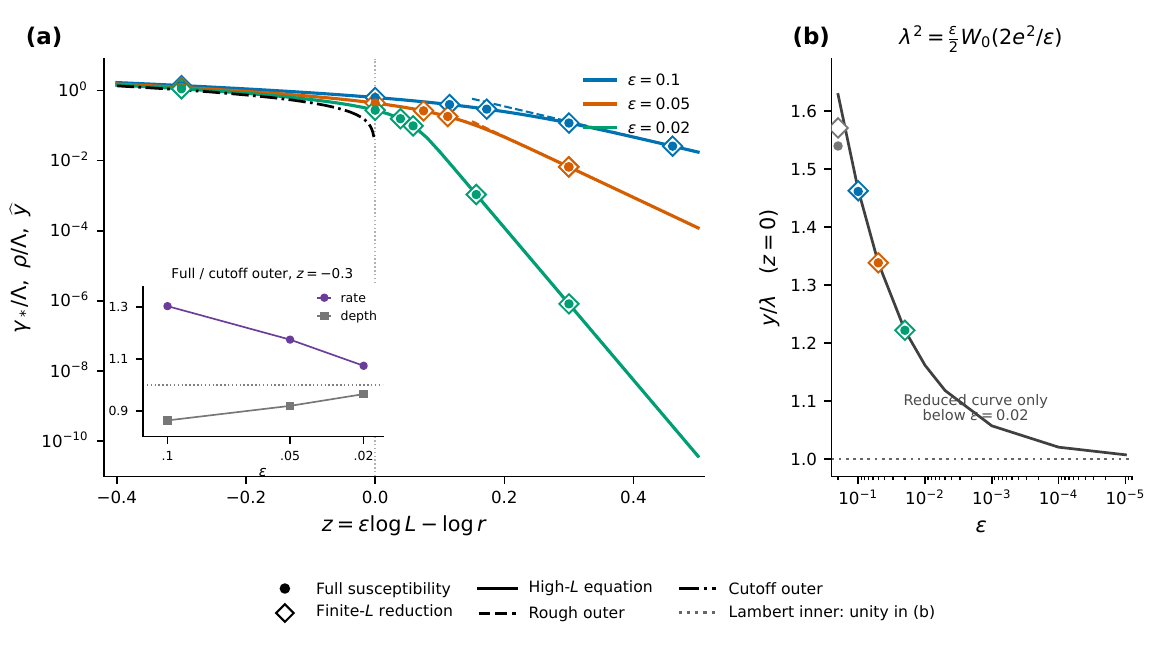}
\caption{Optimizer crossover with numerical and asymptotic provenance separated. In panel (a), filled circles are full-susceptibility optima, open diamonds are independently solved finite-\(L\) covariance roots, and colored solid curves interpolate only the archived high-\(L\) reduced-equation curves at \(\epsilon\) = 0.1, 0.05, and 0.02. The marker positions are not displaced; near-coincident circles and diamonds indicate agreement at the plotting resolution. Colored dashed curves show the rough outer prediction, while the black dash-dotted curve shows the cutoff outer limit. The inset isolates the three full-susceptibility points at \(z\) = -0.3, including the validation point at \(\epsilon\) = 0.02, and divides their optimizer and depth by the respective cutoff outer predictions. Connecting segments in the inset are guides to the eye. Panel (b) tests the Lambert inner scale at \(z\) = 0: the ordinate is \(y/\lambda\), with unity denoting the inner prediction. Filled circles and open diamonds are available at \(\epsilon\) = 0.2, 0.1, 0.05, and 0.02; the gray symbols denote \(\epsilon\) = 0.2. The solid continuation to smaller \(\epsilon\) contains reduced-equation results only. The full-susceptibility ratio remains about 1.22 at \(\epsilon\) = 0.02, making the finite-\(\epsilon\) correction explicit. These finite data support the predicted crossover without replacing the analytic limit arguments}
\label{fig:2}
\end{figure}
\subsection{Different recovery scales for rate and gain}

Recovering the location of the ideal optimum requires a weaker separation from the cutoff than recovering its threshold gain. This distinction is the central result illustrated by Fig. 3. Let \(\gamma_{\rm rough}=\tau^{-1}[(1+\epsilon)r]^{1/\epsilon}\) denote the ideal rough asymptotic reference rate, and define
\begin{equation}
y_{\rm rough}=\frac{\gamma_{\rm rough}}{\Lambda},\qquad
\Delta K_{\rm rough}=\frac{2s\epsilon}{\beta_s(1+\epsilon)\gamma_{\rm rough}}.
\end{equation}
Along the bounded-\(\xi\) paths specified above, Corollary~\ref{recovery} gives
\begin{equation}
\begin{gathered}\frac{\gamma_*}{\gamma_{\rm rough}}\to1
\quad\Longleftrightarrow\quad
\frac{y_{\rm rough}}{\sqrt\epsilon}\to0,
\\
\frac{\Delta K}{\Delta K_{\rm rough}}\to1
\quad\Longleftrightarrow\quad
\frac{y_{\rm rough}}{\epsilon}\to0.\end{gathered}
\end{equation}
These are necessary and sufficient conditions within that class of paths.

For fixed \(\xi\), \(y_{\rm rough}\sim e\epsilon^\xi\). Rate recovery therefore requires \(\xi>1/2\), whereas depth recovery requires \(\xi>1\). The equality cases are explicit: at \(\xi=1/2\), \(\gamma_*/\gamma_{\rm rough}\to e^{-1}\), while at \(\xi=1\), \(\Delta K/\Delta K_{\rm rough}\to1+e\); hence equality does not recover the corresponding ideal quantity. In particular, throughout \(1/2<\xi<1\), the optimal rate approaches its ideal rough value while the threshold gain retains sensitivity to the ultraviolet regularization. Agreement in the optimizer position is consequently insufficient to establish agreement in the optimized gain.

Figure 3 uses only full-susceptibility points from the full-susceptibility validation set. For \(\xi=3/4\), decreasing \(\epsilon\) through \(0.1,0.05,0.02\) raises the rate ratio through approximately \(0.634,0.645,0.681\), while the depth ratio increases through approximately \(3.97,5.23,7.14\). The rate remains appreciably below unity: the data show a trend consistent with eventual rate recovery, whereas the recovery itself is an asymptotic theorem. The simultaneous departure of the depth ratio from unity makes the scale separation visible at finite \(\epsilon\). The \(\xi=2\) series approaches unity in both quantities, while \(\xi=1/2\) lies on the rate-recovery boundary.

\begin{figure}[H]
\centering
\includegraphics[width=\textwidth]{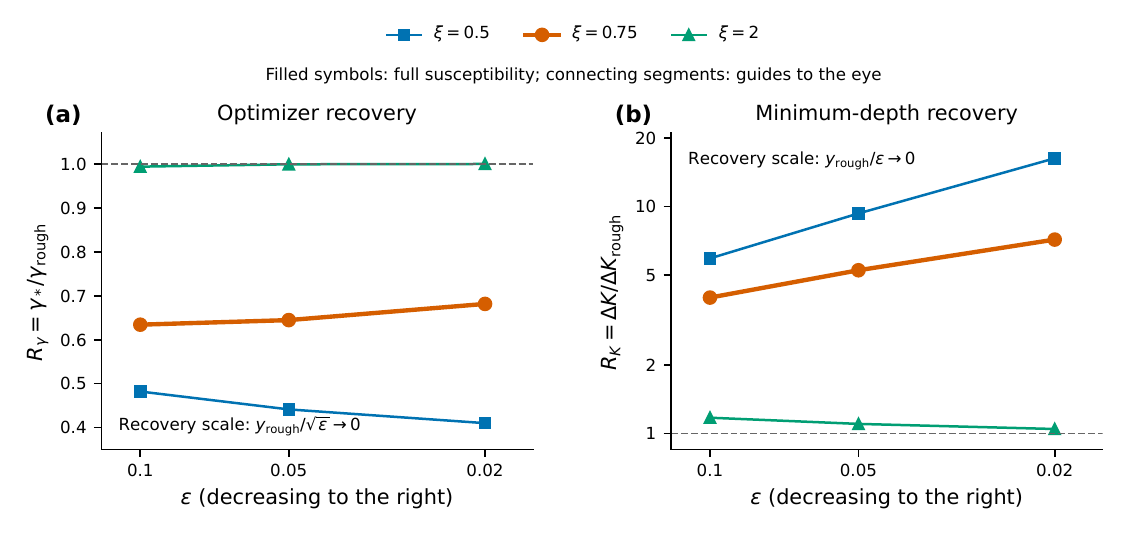}
\caption{Rate and minimum depth recover on different cutoff scales. All filled symbols are full-susceptibility results from the three inner-series subsets of the twelve-point full-susceptibility validation set, with \(\epsilon\) = 0.1, 0.05, and 0.02 and \(z=\xi\epsilon\log(1/\epsilon)\). The three fixed values \(\xi\) = 1/2, 3/4, and 2 are distinguished by color and marker shape; connecting segments are guides to the eye, not fitted or extrapolated curves. Epsilon decreases from left to right. Panel (a) shows the optimizer relative to the rough-model prediction, whereas panel (b) shows the minimum depth relative to its rough-model prediction on a logarithmic vertical scale. Dashed unity lines indicate complete recovery of the corresponding rough-model quantity. The emphasized \(\xi\) = 3/4 series lies between the two recovery boundaries: its rate ratio increases from about 0.634 to 0.681, while its depth ratio increases from about 3.97 to 7.14. Thus the finite-\(\epsilon\) data exhibit the predicted separation of trends, although the rate has not yet reached its limiting value. The \(\xi\) = 2 series approaches unity in both panels, while \(\xi\) = 1/2 does not recover the rate. The displayed scale criteria summarize the established corollary; the three points per series illustrate its consequences without numerically proving an asymptotic limit}
\label{fig:3}
\end{figure}
\subsection{Dependence on cutoff convention}

Figure 4 compares the filtered-white endpoints, \(\epsilon=0\), for the Gaussian and Lorentzian spectral conventions at \(r=2\). The filters are \(W_{\rm G}(\omega)=\exp[-(\omega/\Lambda)^2]\) and \(W_{\rm L}(\omega)=[1+(\omega/\Lambda)^2]^{-1}\), with the same zero-frequency noise intensity. The comparison concerns their large-bandwidth endpoint coefficients for \(\gamma_*/\Lambda\) and \(\Lambda\Delta K\).

For the Gaussian endpoint, \(C_{\rm G}(t)=D\Lambda\pi^{-1/2}e^{-(\Lambda t)^2/4}\), so with \(y=\gamma/\Lambda\) its covariance Laplace transform is \(B_{\rm G}(\gamma)=D b_{\rm G}(y)\), \(b_{\rm G}(y)=e^{y^2}\operatorname{erfc}(y)\). The large-bandwidth stationarity condition is \(b_{\rm G}(y_{\rm G})-y_{\rm G}b'_{\rm G}(y_{\rm G})=1/r\), with \(\gamma_*/\Lambda\to y_{\rm G}\) and \(\Delta K\sim[2D/(\beta_s\Lambda)][-b'_{\rm G}(y_{\rm G})]\). At \(r=2\), \(y_{\rm G}=1.8243589658\ldots\); for \(D=1\), \(\beta_s=2\), this gives \(\Lambda\Delta K\to0.1230391637\ldots\). For the Lorentzian endpoint, \(b_{\rm L}(y)=(1+y)^{-1}\), so the same condition gives \(y_{\rm L}=1+\sqrt2\) and \(\Lambda\Delta K\to(2+\sqrt2)^{-2}=0.0857864376\ldots\). Thus the two references in Fig.~4 follow from the same endpoint optimization rule.

The Lorentzian endpoint has a larger optimizer coefficient and a smaller depth coefficient than the Gaussian endpoint. The plotted Gaussian full-susceptibility points approach the Gaussian coefficients; the Lorentzian references are analytic endpoint values, with no finite-bandwidth Lorentzian full points included in this figure. This endpoint comparison is separate from the interior fixed-\(z\) limit in Theorem~\ref{main}(ii).

The coefficients are compared using the stated spectral definitions of \(\Lambda\); an operational bandwidth match would require specifying that convention. The conclusion is limited to the demonstrated dependence: the UV scale alone does not determine the finite optimizer and depth coefficients.
\begin{figure}[H]
\centering
\includegraphics[width=\textwidth]{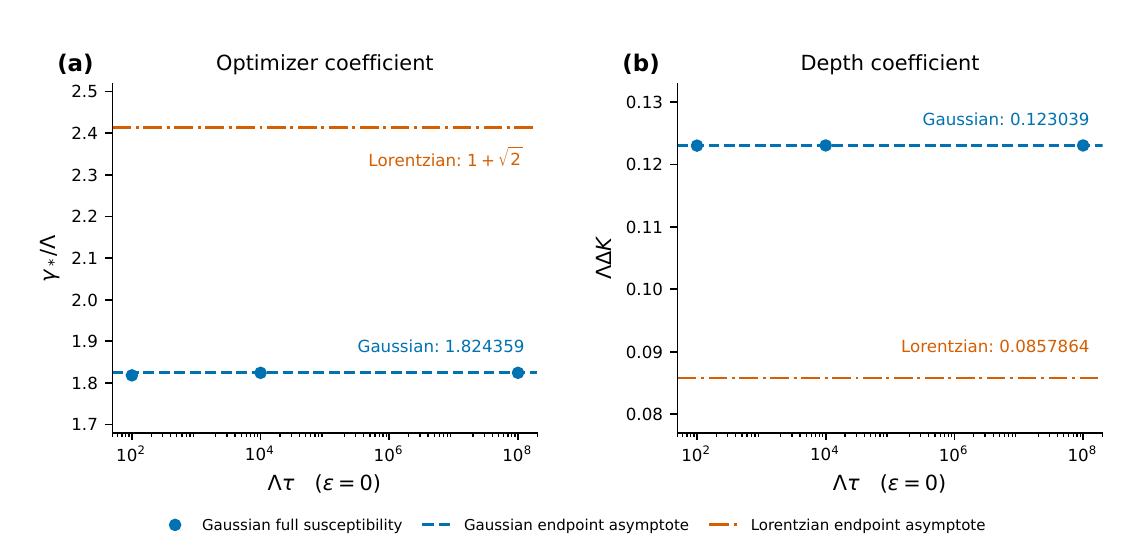}
\caption{Finite optimizer and depth coefficients depend on the cutoff shape. The comparison is restricted to the filtered-white endpoint, \(\epsilon\) = 0, with \(D=\tau=\mu=1\) and \(\beta_s=2\). Panel (a) shows the dimensionless optimizer \(\gamma_*/\Lambda\), and panel (b) shows the scaled threshold depth \(\Lambda\Delta K\). Filled blue circles are the full-susceptibility calculations for the Gaussian spectral filter at \(\Lambda\tau\) = \(10^2\), \(10^4\), and \(10^8\). The blue dashed lines give the Gaussian large-bandwidth endpoint coefficients; orange dash-dotted lines give the corresponding analytic Lorentzian coefficients. No finite-bandwidth Lorentzian full-susceptibility points are represented in this figure. The Gaussian full data approach their reference coefficients, while the two analytic endpoint values remain distinct in both panels. Consequently, the existence of an ultraviolet cutoff alone does not determine these finite coefficients. The comparison uses the stated spectral conventions, \(W_{\rm G}(\omega)=e^{-(\omega/\Lambda)^2}\) and \(W_{\rm L}(\omega)=[1+(\omega/\Lambda)^2]^{-1}\), with the same zero-frequency intensity. Matching a different operational definition of bandwidth could change the numerical comparison. This endpoint figure illustrates shape dependence within this numerical comparison; it neither asserts universality across other regularizations nor adds a finite-\(\epsilon\) Gaussian parameter scan}
\label{fig:4}
\end{figure}
\section{Discussion}
Consistently with the criterion in Ref. \cite{hayashi2026}, existence of an interior optimum in the present model depends on the full phase-correlation structure, while the short-time structure controls the fast-switching asymptotics. Finite ultraviolet regularization determines how the exponentially selected rough-model optimizer is modified. These roles cannot be reduced to a single memory time or to the roughness exponent alone. The cutoff resolves the shortest-time behavior, and its competition with the ideal selection scale becomes singular as \(\epsilon\) decreases.

The origin of the two recovery scales is already visible in the high-bandwidth small-\(y\) expansion. With \(J_\epsilon(y)=y^{-\epsilon}-y+y^{2-\epsilon}+\cdots\), the stationarity combination \(\mathcal T_\epsilon=J_\epsilon-yJ'_\epsilon\) cancels the linear cutoff term \(-y\). The optimizer displacement is therefore controlled first by the competition of \(y^2\) with \(\epsilon\), which leads to the weaker rate-recovery scale \(y_{\rm rough}/\sqrt\epsilon\). By contrast, the depth depends on \(-J'_\epsilon(y)=\epsilon y^{-1-\epsilon}+1+\cdots\): the same linear cutoff correction survives there as the constant term \(1\). Recovering the small ideal depth therefore requires the stronger scale \(y_{\rm rough}/\epsilon\). The rate--gain separation is thus a consequence of how the ultraviolet correction enters the stationarity condition and the optimized objective differently, rather than an arbitrary division into asymptotic regions.

An observed optimal rate may agree with the ideal rough prediction even when the threshold reduction achieved at that rate remains sensitive to short-time noise structure. Locating the minimum and resolving its depth therefore test different aspects of the idealization. In the intermediate range \(1/2<\xi<1\), the theorem makes this distinction explicit: the optimizer position recovers before its gain does. Moreover, the absolute gain can become very small even when its ratio to the rough prediction differs substantially from unity. Here gain refers specifically to the reduction of the linear instability threshold relative to its fast-switching limit.

The linear threshold analysis does not determine the amplitude, stability, or dynamics of the nonlinear ordered state, nor whether a small threshold reduction is observable at finite system size. The full-susceptibility calculations evaluate the linear threshold theory and are not many-body simulations. Likewise, the endpoint comparison uses specified spectral cutoff conventions; it does not establish a mapping between an experimental bandwidth measurement and the model parameter \(\Lambda\). These limitations separate the established rate and depth asymptotics from nonlinear, finite-size, and experimental claims.
\section{Conclusion}
A finite ultraviolet scale modifies the essentially divergent optimal switching rate of the ideal rough model through a singular crossover. For the Lorentzian regularization, the full-susceptibility optimum is controlled across the rough outer, cutoff outer, and Lambert inner regions. The results concern the true global optimizer and the minimum depth with relative error control, rather than only roots of a reduced covariance equation.

The central distinction is that rate recovery precedes gain recovery. Within the bounded inner scaling paths, rate recovery is equivalent to \(y_{\rm rough}/\sqrt\epsilon\to0\), whereas depth recovery is equivalent to \(y_{\rm rough}/\epsilon\to0\). An optimal rate consistent with the ideal description can therefore coexist with a threshold reduction that still depends on ultraviolet regularization.

These results delimit the use of idealized rough or white-noise descriptions in switching optimization. The short-time limit relevant to a rapidly selected optimizer must be resolved together with the noise regularization. Convergence to a white-noise description at fixed switching rate does not by itself determine an optimizer whose rate diverges in the same limit. The endpoint comparison further shows that a nominal UV scale does not fix the limiting scaled coefficients without a cutoff convention. These conclusions are restricted to the stated Gaussian forcing families, asymptotic paths, and mean-field linear instability threshold.

\endgroup
\clearpage
\appendix
\section{Lorentzian cutoff: optimizer and relative-depth asymptotics}

Fix $D,\tau,s,\beta_s>0$ with $r=D/s>1$. We let $\eps\downarrow0$ and
$L=\Lambda\tau\to\infty$, taking $0<\eps\le1/2$ without loss of generality.
With the convention $S(\omega)=\int_{\mathbb R}C(t)e^{-i\omega t}dt$, set
\begin{equation}\label{setup}
 S(\omega)=\frac{2D\operatorname{Re}(1+i\omega\tau)^{-\eps}}
 {1+\omega^2/\Lambda^2},\quad
 Q(t)=\int_0^t(t-u)C(u)du,\quad
 H(\gamma)=(\gamma+s)\int_0^\infty e^{-\gamma t-Q(t)}dt.
\end{equation}
We maximize $H$ over $\gamma\ge0$; equivalently, we minimize
$K_c=2/(\beta_sH)$. Write
$\DD=2/\beta_s-K_c(\gamma_*)$,
$a=L^{-1}$, $q=L^{-\eps}$, $y=\gamma/\Lambda$,
$z=\eps\log L-\log r$, and $h=\log(1/\eps)$; thus $Dq=se^{-z}$.
Constants may depend on these fixed physical parameters and on the stated
compact sets or finite $M,Y$, but not on $\eps,L,z,\xi$ within those sets.

For $p>0$ and $a\ge0$, define the continuous extensions at $y=1$ of
\begin{equation}\label{bdef}
 b_{p,a}(y)=\frac{(y+a)^{-p}-y(1+a)^{-p}}{1-y^2},\quad
 J_p=b_{p,0},\quad t_{p,a}=b_{p,a}-yb'_{p,a},\quad
 \Tr=J_\eps-yJ'_\eps .
\end{equation}
Let $B=\mathcal LC$, $T=B-\gamma B'$ and distinguish the two auxiliary roots
$\Tr(\widehat y)=e^z$ and $T(\rho)=s$. The full optimizer is denoted by $\gamma_*$.

\begin{lemma}[Logarithmic root stability]\label{rootlemma}
Let $f\in C^1$, $f(x_0)=0$, and $-\Logder f\ge kA$ on
$|\log(x/x_0)|\le b$, where $k,A,b>0$. If a continuous perturbation $d$ satisfies
$|d|\le\eta A$ there for $\eta\ge0$ and $2\eta/k<b$, then $f+d$ has a zero in
$|\log(x/x_0)|\le2\eta/k$; every zero in the original neighborhood satisfies
$|\log(x/x_0)|\le\eta/k$. If uniqueness is known separately, this is the
unique relevant root. No derivative bound on $d$ is required.
\end{lemma}

\begin{proposition}[Transfer to the full susceptibility]\label{transfer}
The roots $\widehat y,\rho$ are unique. For $\widehat y\le Y$ and
$a/\widehat y\to0$, with
$\kappa_\eps(y)=y^2J''_\eps(y)/\Tr(y)$ and
$\kappa_B=\rho^2B''(\rho)/s$,
\begin{align}
 0\le\log\frac{\Lambda\widehat y}{\rho}
 &\lesssim\frac{\eps a}{\widehat y(\eps+\widehat y^2)},\qquad
 \frac{\kappa_B}{\kappa_\eps(\widehat y)}=1+O(a/\widehat y).
 \label{finiteroot}
\end{align}
If additionally $\ee=a/[\widehat y(\eps+\widehat y^2)]\to0$, the full
susceptibility has a unique interior global maximizer and
\begin{equation}\label{transferresult}
 \left|\log\frac{\gamma_*}{\Lambda\widehat y}\right|=O(\ee),\qquad
 \DD=\frac{2Dq}{\beta_s\Lambda}[-J'_\eps(\widehat y)]\,[1+O(\ee)].
\end{equation}
\end{proposition}

\begin{theorem}[Three regimes]\label{main}
For the full optimizer and depth in \eqref{setup}, the following hold.
\begin{enumerate}
\item Uniformly for $z$ in any compact subset of $(0,\infty)$,
\begin{equation}\label{rough}
 \gamma_*\tau\sim[(1+\eps)r]^{1/\eps},\qquad
 \DD\sim\frac{2s\eps}{\beta_s(1+\eps)\gamma_*}.
\end{equation}
\item Uniformly for $z$ in any compact subset of $(-\log r,0)$,
\begin{equation}\label{cutoff}
 \gamma_*/\Lambda\to y_0(z):=e^{-z}-1+\sqrt{e^{-z}(e^{-z}-1)},\qquad
 \DD\sim\frac{2s}{\beta_s\Lambda[1+2y_0(z)]}.
\end{equation}
\item For each fixed finite $M$, let $z=\eps h\xi$, $|\xi|\le M$, and
\begin{equation}\label{lambert}
 \lambda^2=\frac\eps2W_0\!\left(\frac2\eps e^{2-2z/\eps}\right).
\end{equation}
Then uniformly on $|\xi|\le M$,
\begin{equation}\label{inner}
 \frac{\gamma_*}{\Lambda\lambda}\to1,\qquad
 \frac{\DD}{[2s/(\beta_s\Lambda)](1+\eps/\lambda)}\to1.
\end{equation}
Both relative errors are bounded by
$C_M[\eps h^2+\sqrt{\eps h}+r^{-1/\eps}\eps^{-2M-2}]$.
At $z=0$, $\gamma_*/\Lambda\sim\sqrt{(\eps/2)\log(1/\eps)}$
and $\DD\sim2s/(\beta_s\Lambda)$.
\end{enumerate}
\end{theorem}

\begin{corollary}[Distinct recovery scales]\label{recovery}
Set $\gr=\tau^{-1}[(1+\eps)r]^{1/\eps}$, $\yr=\gr/\Lambda$, and
$\Dr=2s\eps/[\beta_s(1+\eps)\gr]$.
Along every path in the bounded-$\xi$ domain of Theorem~\ref{main},
\begin{equation}\label{iff}
 \frac{\gamma_*}{\gr}\to1\iff\frac{\yr}{\sqrt\eps}\to0,
 \qquad
 \frac{\DD}{\Dr}\to1\iff\frac{\yr}{\eps}\to0.
\end{equation}
For fixed $\xi$, these are respectively $\xi>1/2$ and $\xi>1$;
the equality cases do not recover the ideal asymptotics.
\end{corollary}

\paragraph{1. Exact identities and global uniqueness.}
The covariance is the convolution of $w_\Lambda(t)=(\Lambda/2)e^{-\Lambda|t|}$
and $C_\eps(t)=D|t|^{\eps-1}e^{-|t|/\tau}/[\tau^\eps\Gamma(\eps)]$.
Thus $C$ is positive, bounded, even, and nonincreasing on $[0,\infty)$;
the last assertion follows by representing even decreasing functions as
positive mixtures of symmetric-interval indicators.
Moreover $\int_0^\infty C=D$, $C(0)=D\Lambda(1+L)^{-\eps}$, and
$(1-\Lambda^{-2}\partial_t^2)C=C_\eps$, with $C'(0)=0$, gives
\begin{equation}\label{Bexact}
 B(\gamma)=D\frac{(1+\gamma\tau)^{-\eps}
 -(\gamma/\Lambda)(1+L)^{-\eps}}{1-(\gamma/\Lambda)^2}
 =Dq\,b_{\eps,a}(y).
\end{equation}
In particular $m_1=-B'(0)=D\eps\tau+D(1+L)^{-\eps}/\Lambda$ and
$0\le Dt-Q(t)\le m_1$, so all phase moments exist and
$e^{-Dt}\le e^{-Q(t)}\le e^{m_1-Dt}$.
Writing $A(t)=Q'(t)$, integration by parts yields
$H'=\int_0^\infty t e^{-\gamma t}[A(t)-s]e^{-Q(t)}dt$.
Since $A$ increases strictly from $0$ to $D$, its integrand changes sign
once, from negative to positive. At a zero of $H'$, multiplying by
$e^{-vt}-e^{-vt_s}$, where $A(t_s)=s$, shows that $H'$ is negative at
every larger $\gamma$ and positive at every smaller $\gamma\ge0$.
Hence any interior critical point is the unique global maximum.
Exactly, it exists iff $M_0>sM_1$, where $M_j=\int_0^\infty t^je^{-Q(t)}dt$;
otherwise the maximum is at zero. Here
$H'(0)\ge(D-se^{m_1})/D^2>0$ eventually, while
$H'=-s/\gamma^2+O(\gamma^{-3})<0$ at large $\gamma$ for each fixed model.

\paragraph{2. One root lemma, two perturbations, and relative depth.}
For Lemma~\ref{rootlemma}, integrate $-\Logder f\ge kA$ from $x_0$.
At $x_0e^{\pm2\eta/k}$ the unperturbed signs dominate $d$, giving a zero
by continuity. At any zero, $|f|=|d|\le\eta A$ gives the sharper bound.
This proves the lemma without assuming monotonicity of the perturbed function.

The apparent pole in \eqref{bdef} is removed by the positive identity
\begin{equation}\label{regular}
 b_{p,a}(y)=\frac{(1+a)^{-p}
 +p\int_0^1(1-u+uy+a)^{-p-1}du}{1+y}.
\end{equation}
Differentiation shows complete monotonicity. Since
$\partial_a b_{p,a}=-p b_{p+1,a}$, uniformly on $0<y\le Y$, $k=0,1,2$,
\begin{equation}\label{derror}
 0\le(-1)^k(J_\eps^{(k)}-b_{\eps,a}^{(k)})
 \le C\eps a y^{-\eps-k-1},\qquad
 \frac{(-1)^kb_{p,a}^{(k)}}{(-1)^kJ_p^{(k)}}
 \in[(1+a/\min(1,y))^{-p-k-1},1].
\end{equation}
Indeed the first bound follows by integrating $\partial_a b$; for
$y\le1/2$ differentiate the quotient, and for $y\ge1/2$ use
\eqref{regular} and compactness. The second follows termwise from its
positive Leibniz expansion. That expansion also bounds the logarithmic
decay rate of $(-1)^kb_{p,a}^{(k)}$ by $p+k+2$.

The convergent series $J_\eps=\sum_{n\ge0}y^{2n-\eps}-\sum_{n\ge0}y^{2n+1}$
near zero, together with \eqref{regular} away from zero, gives
\begin{equation}\label{slope}
 \Tr\asymp y^{-\eps},\qquad
 J''_\eps\asymp\eps y^{-\eps-2}+y^{-\eps},\qquad
 \kappa_\eps\asymp\eps+y^2\quad(0<y\le Y).
\end{equation}
For the lower bound, the positive terms
$\eps(1+\eps)y^{-\eps-2}$ and $(2-\eps)(1-\eps)y^{-\eps}$
absorb the negative $O(y)$ remainder; away from zero,
$J_0=(1+y)^{-1}$ and $J_0''>0$ give a uniform bound.
Since $t'_{p,a}=-yb''_{p,a}<0$, its endpoint values prove uniqueness of
both auxiliary roots. Equations~\eqref{derror}--\eqref{slope} imply
$0\le\Tr-t_{\eps,a}\lesssim\eps a y^{-\eps-1}$.
Apply Lemma~\ref{rootlemma} on a fixed logarithmic neighborhood of
$\widehat y$ with $A=e^z$, $k\asymp\eps+\widehat y^2$ and
$\eta\lesssim\eps a/\widehat y$. This proves the first part of
\eqref{finiteroot}. The derivative ratios in \eqref{derror} and bounded
logarithmic decay rates then give its second part and
\begin{equation}\label{depthfinite}
 -B'(\rho)=\frac{Dq}{\Lambda}[-J'_\eps(\widehat y)]
 [1+O(a/\widehat y)].
\end{equation}

For the second perturbation, $0\le Q\le Dt$ and
$0\le e^{-Q}-1+Q\le Q^2/2$ imply
\begin{equation}\label{Herror}
 \begin{gathered}
 \mathcal L(e^{-Q})=\gamma^{-1}-B\gamma^{-2}+R,
 \quad 0\le R\le D^2\gamma^{-3},\quad |R'|\le3D^2\gamma^{-4};\\
 \gamma^2H'=-s+T+O(\gamma^{-1}).
 \end{gathered}
\end{equation}
The constants are uniform for $\gamma\ge\max(D,s)$.
Put $I_k=\int_0^\infty t^ke^{-\gamma t}C(t)dt$.
The covariance inequality between increasing $t$ and decreasing $C(t)$
under a Gamma$(k+1,\gamma)$ law gives $\gamma I_{k+1}/I_k\le k+1$.
Consequently
\begin{equation}\label{motion}
 \frac{\gamma B''}{-B'}\le2,\quad
 -1\le\partial_{\log\gamma}\log(\gamma^2B'')\le2,\quad
 \left|\log\frac{-B'(\gamma_2)}{-B'(\gamma_1)}\right|
 \le2\left|\log\frac{\gamma_2}{\gamma_1}\right|.
\end{equation}
On a fixed logarithmic neighborhood of $\rho$, apply the same lemma to
$f=T-s$, $A=s$, $k\asymp\kappa_B$, and the error in \eqref{Herror}.
If $\rho\kappa_B\to\infty$, global uniqueness identifies its root and
$|\log(\gamma_*/\rho)|=O((\rho\kappa_B)^{-1})$.
This division by the small slope is essential: on the rough side
$\kappa_B\sim\eps$, so an unspecified $o(1)$ error would not suffice.

At the true root, \eqref{Herror} gives
$H(\gamma_*)-1=-B'(\gamma_*)+O(\gamma_*^{-2})$.
The needed \emph{relative}, rather than additive, control is
\begin{equation}\label{relative}
 \frac{(sD+D^2)/\rho^2}{-B'(\rho)}
 \le\frac{2(sD+D^2)}{s\rho\kappa_B}\to0.
\end{equation}
Together with \eqref{motion}, this controls both the remainder and root
movement. Since $\kappa_B\le2$ and $-B'(\rho)\le D/\rho$,
$\DD=(2/\beta_s)[-B'(\rho)]
[1+O((\rho\kappa_B)^{-1})]$.
Combining \eqref{finiteroot}, \eqref{slope}, \eqref{depthfinite}, and
$\Lambda^{-1}=\tau a$ proves Proposition~\ref{transfer}.

\paragraph{3. Local Lorentzian asymptotics.}
For $v=y^{-\eps}$ and $y\le1/2$, the exact remainder
$J_\eps=v-y+y^2v-y^3+R_J$,
$R_J=(y^4v-y^5)/(1-y^2)$ satisfies
$|\partial_y^kR_J|\lesssim y^{4-k}v+y^{5-k}$, $k\le2$.
Thus, without differentiating unspecified little-$o$ terms,
\begin{align}
 \Tr&=(1+\eps)v-(1-\eps)y^2v+2y^3+O(y^4v+y^5),\notag\\
 -J'_\eps&=\eps v/y+1-(2-\eps)yv+3y^2+O(y^3v+y^4),\label{local}\\
 y^2J''_\eps&=\eps(1+\eps)v+(2-\eps)(1-\eps)y^2v-6y^3
 +O(y^4v+y^5).\notag
\end{align}

\paragraph{4. Outer and inner regimes.}
For $z\in I_+\Subset(0,\infty)$, let
$\yr=[(1+\eps)e^{-z}]^{1/\eps}$.
Equation~\eqref{local} and Lemma~\ref{rootlemma} give
$\log(\widehat y/\yr)=O(\yr^2/\eps)=o(1)$,
$\kappa_\eps(\widehat y)\sim\eps$, and
$-J'_\eps(\widehat y)\sim\eps\widehat y^{-1-\eps}$.
Here $L\yr=[(1+\eps)r]^{1/\eps}$ and
$\ee=O(r^{-1/\eps}/\eps)\to0$ uniformly.
Substitution in \eqref{transferresult} proves \eqref{rough}.
For $z\in I_-\Subset(-\log r,0)$, compact convergence to
$\mathcal T_0=(1+2y)/(1+y)^2$ gives
$\widehat y\to y_0$ and
$\kappa_\eps(\widehat y)\to2y_0^2/[(1+y_0)(1+2y_0)]>0$.
Since $a\to0$ exponentially and $Dq=se^{-z}$,
\eqref{transferresult} proves \eqref{cutoff}, including intervals crossing $y=1$.

In the inner domain, $\lambda$ in \eqref{lambert} is the zero of
$f(y)=\eps(1-\log y)-z-y^2$.
Evaluating $f$ at $\eps^{M+1}$ and $\sqrt{(M+2)\eps h}$ gives
$\eps^{M+1}<\lambda\le\sqrt{(M+2)\eps h}$.
On any fixed logarithmic neighborhood of $\lambda$,
$|\log y|\lesssim_M h$ and \eqref{local} yields
\begin{equation}\label{residual}
 \Tr(y)-e^z=f(y)+R_\eps(y),\qquad
 |R_\eps(y)|\le C_M\{\eps^2h^2+\eps h y^2+y^3\}.
\end{equation}
The logarithmic slope of $f$ is $\eps+2y^2\asymp\eps+\lambda^2$;
dividing \eqref{residual} by this scale gives
$O_M(\omega_\eps)$, where
$\omega_\eps=\eps h^2+\sqrt{\eps h}\to0$.
Lemma~\ref{rootlemma} therefore gives
$|\log(\widehat y/\lambda)|=O_M(\omega_\eps)$ uniformly.
Equation~\eqref{local} also gives, on this neighborhood,
\begin{equation}\label{innerlocal}
 \kappa_\eps(y)=(\eps+2y^2)[1+O_M(\eps h+y)],\qquad
 -J'_\eps(y)=(1+\eps/y)[1+O_M(\eps h+y)].
\end{equation}
Since $a=r^{-1/\eps}\eps^\xi$ and $\widehat y\sim\lambda$,
$\ee\lesssim_M r^{-1/\eps}\eps^{-2M-2}\to0$.
Use \eqref{transferresult}, \eqref{innerlocal} and $Dq=se^{-z}\sim s$
to obtain \eqref{inner} and its stated quantitative error.
Thus both finite-$L$ and full-susceptibility errors vanish even after
division by the shrinking slope. In particular, the depth remainder
divided by depth is respectively
$O((\eps\rho)^{-1})$, $O(\Lambda^{-1})$, and
$O_M([\Lambda\lambda(\lambda+\eps)]^{-1})$ in the three regimes;
all vanish. The specialization at $z=0$ follows from
$W_0(2e^2/\eps)\sim\log(1/\eps)$.

\paragraph{5. Rate--depth scale separation.}
In the inner domain put $Y=\widehat y$. The root equation and
\eqref{local} reduce the proof to two identities:
\begin{align}
 \log(\yr/Y)&=\frac{Y^2}{\eps}[1+O(\eps+Y)],\label{rateid}\\
 \frac{-J'_\eps(Y)}{\eps\yr^{-1-\eps}}
 &=\left(\frac{\yr}{Y}\right)^{1+\eps}
 +\frac{\yr^{1+\eps}}\eps[1+O_M(Y)].\label{depthid}
\end{align}
Indeed $\Tr/[(1+\eps)y^{-\eps}]
=1-[(1-\eps)/(1+\eps)]y^2+O(y^3)$ proves \eqref{rateid};
$Y^{-\eps}=1+O_M(\eps h)$ proves \eqref{depthid}.
Eventually $Y\le\yr$, so \eqref{rateid} implies
$Y/\yr\to1$ iff $\yr^2/\eps\to0$.
In \eqref{depthid} both terms are positive, the first is at least one,
and $\yr^\eps\to1$ uniformly. Its left side tends to one iff
$\yr/\eps\to0$; sufficiency also uses the first equivalence.
Finally $\Dr=(2Dq/\beta_s\Lambda)\eps\yr^{-1-\eps}$ and the
\emph{relative} transfer \eqref{transferresult} give both equivalences
in \eqref{iff}, even along paths where the depth ratio diverges.
For fixed $\xi$, $\yr\sim e\eps^\xi$, proving the stated strict boundaries.
\hfill$\square$

\end{document}